\input harvmac
\noblackbox

\input epsf

\def\hat{\widehat}
\newcount\figno
\figno=0
\def\fig#1#2#3{
\par\begingroup\parindent=0pt\leftskip=1cm\rightskip=1cm\parindent=0pt
\baselineskip=11pt
\global\advance\figno by 1
\midinsert
\epsfxsize=#3
\centerline{\epsfbox{#2}}
\vskip 12pt
{\bf Fig.\ \the\figno: } #1\par
\endinsert\endgroup\par
}
\def\figlabel#1{\xdef#1{\the\figno}}
\def\encadremath#1{\vbox{\hrule\hbox{\vrule\kern8pt\vbox{\kern8pt
\hbox{$\displaystyle #1$}\kern8pt}
\kern8pt\vrule}\hrule}}

 \def\apm{{\alpha^{\prime}}}


 \font\cmss=cmss10
 \font\cmsss=cmss10 at 7pt
 \def\rlx{\relax\leavevmode}
 \def\inbar{\vrule height1.5ex width.4pt depth0pt}
 \def\IC{\relax\,\hbox{$\inbar\kern-.3em{\rm C}$}}
 \def\IN{\relax{\rm I\kern-.18em N}}
 \def\IP{\relax{\rm I\kern-.18em P}}
 \def\ZZ{\rlx\leavevmode\ifmmode\mathchoice{\hbox{\cmss Z\kern-.4em Z}}
  {\hbox{\cmss Z\kern-.4em Z}}{\lower.9pt\hbox{\cmsss Z\kern-.36em Z}}
  {\lower1.2pt\hbox{\cmsss Z\kern-.36em Z}}\else{\cmss Z\kern-.4em
  Z}\fi}
 \def\IZ{\relax\ifmmode\mathchoice
 {\hbox{\cmss Z\kern-.4em Z}}{\hbox{\cmss Z\kern-.4em Z}}
 {\lower.9pt\hbox{\cmsss Z\kern-.4em Z}}
 {\lower1.2pt\hbox{\cmsss Z\kern-.4em Z}}\else{\cmss Z\kern-.4em
 Z}\fi}
 \def\IZ{\relax\ifmmode\mathchoice
 {\hbox{\cmss Z\kern-.4em Z}}{\hbox{\cmss Z\kern-.4em Z}}
 {\lower.9pt\hbox{\cmsss Z\kern-.4em Z}}
 {\lower1.2pt\hbox{\cmsss Z\kern-.4em Z}}\else{\cmss Z\kern-.4em Z}\fi}

 \def\narrowplus{\kern -.04truein + \kern -.03truein}
 \def\narrowminus{- \kern -.04truein}
 \def\narrowminussub{\kern -.02truein - \kern -.01truein}
 
 \def\half{{1\over 2}}
 
 \def\m{{\mu}}
 \def\n{{\nu}}
 \def\ep{{\epsilon}}
 \def\d{{\delta}}
 \def\t{{\theta}}
 \def\a{{\alpha}}
 
 \def\frac#1#2{{#1\over #2}}
 
 \def\G{{\Gamma}}
 
 \def\g{{\gamma}}
 \def\s{{\sigma}}
 
 \def\b{{\beta}}

 \def\CO{{\cal O}}

 \def\CN{{\cal N}}
 \def\p{\partial}

 \def\apm{{\a^{\prime}}}
 \def\r{\rightarrow}

 \def\IZ{\relax\ifmmode\mathchoice
 {\hbox{\cmss Z\kern-.4em Z}}{\hbox{\cmss Z\kern-.4em Z}}
 {\lower.9pt\hbox{\cmsss Z\kern-.4em Z}}
 {\lower1.2pt\hbox{\cmsss Z\kern-.4em Z}}\else{\cmss Z\kern-.4em Z}\fi}
 \def\IB{\relax{\rm I\kern-.18em B}}
 \def\IC{{\relax\hbox{$\inbar\kern-.3em{\rm C}$}}}
 \def\ID{\relax{\rm I\kern-.18em D}}
 \def\IE{\relax{\rm I\kern-.18em E}}
 \def\IF{\relax{\rm I\kern-.18em F}}
 \def\IG{\relax\hbox{$\inbar\kern-.3em{\rm G}$}}
 \def\IGa{\relax\hbox{${\rm I}\kern-.18em\Gamma$}}
 \def\IH{\relax{\rm I\kern-.18em H}}
 \def\II{\relax{\rm I\kern-.18em I}}
 \def\IK{\relax{\rm I\kern-.18em K}}
 \def\IP{\relax{\rm I\kern-.18em P}}

 \font\cmss=cmss10 \font\cmsss=cmss10 at 7pt
 \def\IR{\relax{\rm I\kern-.18em R}}

 %

 %
 %
 \def\eqnn#1{\xdef #1{(\secsym\the\meqno)}\writedef{#1\leftbracket#1}%
 \global\advance\meqno by1\wrlabeL#1}
 \def\eqna#1{\xdef #1##1{\hbox{$(\secsym\the\meqno##1)$}}
 \writedef{#1\numbersign1\leftbracket#1{\numbersign1}}%
 \global\advance\meqno by1\wrlabeL{#1$\{\}$}}
 \def\eqn#1#2{\xdef #1{(\secsym\the\meqno)}\writedef{#1\leftbracket#1}%
 \global\advance\meqno by1$$#2\eqno#1\eqlabeL#1$$}

\lref\BergmanKZ{
O.~Bergman and M.~R.~Gaberdiel,
``Non-BPS Dirichlet branes,''
Class.\ Quant.\ Grav.\  {\bf 17}, 961 (2000)
[hep-th/9908126].
}

\lref\BergmanRF{
O.~Bergman and M.~R.~Gaberdiel,
``A non-supersymmetric open-string theory and S-duality,''
Nucl.\ Phys.\ B {\bf 499}, 183 (1997)
[hep-th/9701137].
}

\lref\CostaNW{
M.~S.~Costa and M.~Gutperle,
``The Kaluza-Klein Melvin solution in M-theory,''
JHEP {\bf 0103}, 027 (2001)
[hep-th/0012072].
}

\lref\DiaconescuRK{
D.~Diaconescu,
``D-branes, monopoles and Nahm equations,''
Nucl.\ Phys.\ B {\bf 503}, 220 (1997)
[hep-th/9608163].
}

\lref\JackiwXX{
R.~Jackiw and C.~Rebbi,
``Spin From Isospin In A Gauge Theory,''
Phys.\ Rev.\ Lett.\  {\bf 36}, 1116 (1976).
}

\lref\HasenfratzGR{
P.~Hasenfratz and G.~'t Hooft,
``A Fermion - Boson Puzzle In A Gauge Theory,''
Phys.\ Rev.\ Lett.\  {\bf 36}, 1119 (1976).
}

\lref\GutperleMB{
M.~Gutperle and A.~Strominger,
``Fluxbranes in string theory,''
JHEP {\bf 0106}, 035 (2001)
[hep-th/0104136].
}
\lref\GoldhaberDP{
A.~S.~Goldhaber,
``Spin And Statistics Connection For Charge - Monopole Composites,''
Phys.\ Rev.\ Lett.\  {\bf 36}, 1122 (1976).
}

\lref\HashimotoQH{
K.~Hashimoto,
``String junction from worldsheet gauge theory,''
Prog.\ Theor.\ Phys.\  {\bf 101}, 1353 (1999)
[hep-th/9808185].
}

\lref\LeeNV{
K.~Lee and P.~Yi,
``Dyons in N = 4 supersymmetric theories and three-pronged strings,''
Phys.\ Rev.\ D {\bf 58}, 066005 (1998)
[hep-th/9804174].
}
\lref\abs{
M.~Atiyah, R. Bott and A. Shapiro, `` Clifford Modules'', 
Topology { \bf 3}, Suppl. 1 (1964) 3.}

\lref\WittenCD{
E.~Witten,
``D-branes and K-theory,''
JHEP {\bf 9812}, 019 (1998)
[hep-th/9810188].
}
\lref\SenTT{
A.~Sen,
``SO(32) spinors of type I and other solitons on brane-antibrane
pair,''
JHEP {\bf 9809}, 023 (1998)
[hep-th/9808141].
}

\lref\BilloBG{
M.~Billo, B.~Craps and F.~Roose,
``Anomalous couplings of type 0 D-branes,''
hep-th/9908030.
}

\lref\SuyamaBN{
T.~Suyama,
``Closed string tachyons in non-supersymmetric heterotic theories,''
hep-th/0106079.
}

\lref\BergmanRF{
O.~Bergman and M.~R.~Gaberdiel,
``A non-supersymmetric open-string theory and S-duality,''
Nucl.\ Phys.\ B {\bf 499}, 183 (1997)
[hep-th/9701137].
}

\lref\BergmanKM{
O.~Bergman and M.~R.~Gaberdiel,
``Dualities of type 0 strings,''
JHEP {\bf 9907}, 022 (1999)
[hep-th/9906055].
}

\lref\tHooftFV{
G. ~'t Hooft,
``Computation Of The Quantum Effects Due To A Four-Dimensional
Pseudoparticle,''
Phys.\ Rev.\ D {\bf 14}, 3432 (1976)
[Erratum-ibid.\ D {\bf 18}, 2199 (1976)].
}

\lref\GauntlettCW{
J.~P.~Gauntlett and D.~A.~Lowe,
``Dyons and S-Duality in N=4 Supersymmetric Gauge Theory,''
Nucl.\ Phys.\ B {\bf 472}, 194 (1996)
[hep-th/9601085].
}

\lref\ThompsonRW{
D.~M.~Thompson,
``Descent relations in type 0A / 0B,''
hep-th/0105314.
}

\lref\EnglertBU{
F.~Englert, L.~Houart and A.~Taormina,
``Brane fusion in the bosonic string and the emergence of fermionic  strings,''
hep-th/0106235.
}

\lref\CasherRA{
A.~Casher, F.~Englert, H.~Nicolai and A.~Taormina,
``Consistent Superstrings As Solutions Of The D = 26 Bosonic String Theory,''
Phys.\ Lett.\ B {\bf 162}, 121 (1985).
}

\lref\WeinbergZT{
E.~J.~Weinberg,
``Fundamental Monopoles And Multi - Monopole Solutions For Arbitrary Simple Gauge Groups,''
Nucl.\ Phys.\ B {\bf 167}, 500 (1980).
}

\lref\Saha{
M.~N.~Saha, 
``On The Origin Of Mass In Neutrons And Protons,'' 
Indian J.\ Phys.\ {\bf 10}, 145 (1936); 
``Note On Dirac's Theory Of Magnetic Poles,'' Phys.\ Rev.\ {\bf
75} 1968 (1949). 
}

\lref\SuyamaNE{
T.~Suyama,
``Melvin Background in Heterotic Theories,''
hep-th/0107116.
}

\lref\StrathdeeJR{
J.~Strathdee,
``Extended Poincare Supersymmetry,''
Int.\ J.\ Mod.\ Phys.\ A {\bf 2}, 273 (1987).
}

\lref\HashimotoNJ{
K.~Hashimoto, H.~Hata and N.~Sasakura,
``Multi-pronged strings and BPS saturated solutions in SU(N)  supersymmetric Yang-Mills theory,''
Nucl.\ Phys.\ B {\bf 535}, 83 (1998)
[hep-th/9804164].
}
 
\lref\HashimotoZS{
K.~Hashimoto, H.~Hata and N.~Sasakura,
``3-string junction and BPS saturated solutions in SU(3) supersymmetric  Yang-Mills theory,''
Phys.\ Lett.\ B {\bf 431}, 303 (1998)
[hep-th/9803127].
}

\lref\ArgyresPV{
P.~C.~Argyres and K.~Narayan,
``String webs from field theory,''
JHEP {\bf 0103}, 047 (2001)
[hep-th/0101114].
}

\lref\KolTW{
B.~Kol and M.~Kroyter,
``On the spatial structure of monopoles,''
hep-th/0002118.
}

\lref\GauntlettXZ{
J.~P.~Gauntlett, C.~Koehl, D.~Mateos, P.~K.~Townsend and M.~Zamaklar,
``Finite energy Dirac-Born-Infeld monopoles and string junctions,''
Phys.\ Rev.\ D {\bf 60}, 045004 (1999)
[hep-th/9903156].
}

\lref\RitzJK{
A.~Ritz and A.~Vainshtein,
``Long range forces and supersymmetric bound states,''
hep-th/0102121.
}

\lref\BergmanGS{
O.~Bergman and B.~Kol,
``String webs and 1/4 BPS monopoles,''
Nucl.\ Phys.\ B {\bf 536}, 149 (1998)
[hep-th/9804160].
}

\lref\KlebanovYY{
I.~R.~Klebanov and A.~A.~Tseytlin,
``D-branes and dual gauge theories in type 0 strings,''
Nucl.\ Phys.\ B {\bf 546}, 155 (1999)
[hep-th/9811035].
}

\lref\KlebanovCH{
I.~R.~Klebanov and A.~A.~Tseytlin,
``A non-supersymmetric large N CFT from type 0 string theory,''
JHEP {\bf 9903}, 015 (1999)
[hep-th/9901101].
}

\lref\CostaQX{
M.~S.~Costa,
``Intersecting D-branes and black holes in type 0 string theory,''
JHEP {\bf 9904}, 016 (1999)
[hep-th/9903128].
}

\Title
 {\vbox{
 \baselineskip12pt
 \hbox{hep-th/0107165}\hbox{}\hbox{}
}}
 {\vbox{
 \centerline{Fermions in  Bosonic String Theories }
 \centerline{}
 }}

\centerline{ Justin R.  David,
\footnote{$^1$}{justin@vulcan.physics.ucsb.edu} 
Shiraz  Minwalla \footnote{$^2$}{minwalla@pascal.harvard.edu} 
and  Carlos  N\'u\~nez \footnote{$^3$}{nunez@lorentz.harvard.edu}
}
\bigskip
\centerline{ {$^1$}Department of Physics,} 
\centerline{ University of California, Santa Barbara, CA 93106.}
\smallskip
\centerline{ {$^{2,3}$}Jefferson Physical Laboratory,}
\centerline{Harvard University, Cambridge, MA 02138.}
\smallskip

 \vskip .3in 

 \centerline{\bf Abstract}

{We generalize the Jackiw-Rebbi-Hasenfratz-'t Hooft construction of
fermions from bosons to demonstrate the fermionic nature of certain bound states 
involving $SU(N)$ instantons 
in even spatial dimensions and $SO(N)$ instantons in $8k+1$ spatial
dimensions. We use this result to identify several fermionic excitations in
various perturbatively bosonic string theories. 
In some examples we are able to identify these fermions  as  
excitations in known conformal field theories and independently confirm their
fermionic nature. Examples of the fermions we find include certain
3-string junctions in type 0B theory, excitations 
of the 0-p system in type 0A theory, excitations of the stable D-particle
of type O theory, and a rich spectrum of fermions in the bosonic string
compactified on the $SO(32)$ group lattice.}

\vskip 0.5cm
\Date{July 2001}
\listtoc
\writetoc

\newsec{Introduction}

Over the last few years the various apparently distinct superstring theories
have been identified as different corners on the moduli space of a single 
theory. However the relationship of 
the bosonic string, perhaps the simplest string theory, to M theory is
still unclear. 

The bosonic string is qualitatively different from its supersymmetric 
cousins in two important respects; it is tachyonic and it does not contain space time 
fermions in its perturbative spectrum.
However, both these features are also true of the type 0A string which, 
nonetheless, has recently found its conjectural position in the 
M theory family. 
Type 0A theory has been interpreted as type IIA theory with a background RR 2-form 
flux \refs{ \CostaNW} (see also \GutperleMB) or, equivalently, M theory 
compactified on a thermal circle \refs{\BergmanKM, \CostaNW}. 
The type 0A tachyon is conjectured to signal
the instability for this state to decay into 
supersymmetric Type IIA vacuum \refs{\GutperleMB, \CostaNW} . 
These conjectures\foot{Similar 
conjectures have also recently been made relating nonsupersymmetric 
heterotic string theories to their supersymmetric counterparts \refs{
\SuyamaBN, \SuyamaNE}.}
are consistent 
 with the purely bosonic nature of perturbative type 0A theory as
the  massless fermions of M-theory are driven out of the perturbative 
type 0A spectrum by the anti-periodic boundary conditions on 
the M-theory circle.
However they reappear in the non-perturbative type 0A
spectrum with energies of order ${1\over R_{11}}={1\over g}$ as excitations
on D-branes  \BergmanKM.  

It is natural to wonder if other perturbatively  
bosonic string theories, 
including the bosonic string,  are connected to the 
M-theory mainland in a manner analogous to that conjectured for the 
0A theory. In this paper we begin an investigation of this question. 
In particular we identify a spectrum of non-perturbative fermionic 
excitations in several  bosonic string theories, including the bosonic string.

Over twenty five years ago Jackiw and Rebbi \JackiwXX\  Hasenfratz and 't Hooft 
\HasenfratzGR\ argued that certain purely bosonic $SU(2)$ 
gauge
theories possess nonperturbative fermionic excitations. As string
theories often contain gauge fields, generalizations 
of the Jackiw -Rebbi -Hasenfratz-'t Hooft construction allow  
us to identify several nonperturbative fermions in perturbatively bosonic 
string theories.

As we review in Section 2, Jackiw, Rebbi, Hasenfratz and 't Hooft
argue that the bound state of an $SU(2)$ 't Hooft-Polyakov monopole
with a quantum that transforms in the fundamental of $SU(2)$, is 
fermionic. It is rather amusing to note (see Section 3 for details) 
that the familiar 3-string junction sketched in Fig 1.
\fig{A 3-string junction. The fundamental string emerging
from D3-brane M meets a D-string emerging from the D3-brane A
to form a (1,1) string which ends on the D3-brane B. The D3-branes 
are parallel, but are separated in transverse space.}{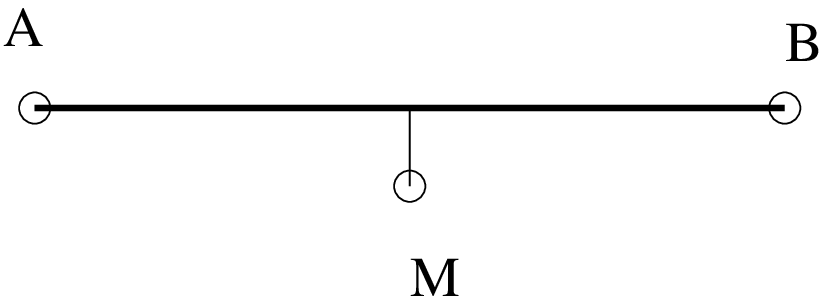}
{4.0truein} \noindent
is a precise implementation of the Jackiw -Rebbi -Hasenfratz - 't Hooft
construction on the world volume of D3-branes in bosonic type 0B string theory 
and perhaps also in bosonic string theory. Consequently, 
the 3-string junction of Fig. 1. is a fermion!

In Section 4 of this paper we generalize the Jackiw-Rebbi-Hasenfratz-'t
Hooft construction.  We argue that, for sufficiently large $N$, 
that bound states involving instantons of $SU(N)$ gauge theories in even
spatial dimensions,
and $SO(N)$ gauge theories in $8k+1$ spatial dimensions, 
are fermionic. In particular, we demonstrate that bound states of 
the familiar $SU(2)$ instanton with a fundamental quantum is 
a fermionic particle in $4+1$ dimensional $SU(N)$ gauge theory. 

In section 5, we employ these constructions to identify several 
fermions in peturbatively bosonic string theories. 
The fermions we study fall into two categories. 
Excitations of the first sort propagate on D-branes, and are bound states 
of perturbative quanta with 
nonperturbative lumps constructed out of D-brane gauge fields. In some 
of the examples that we study, 
 it is possible to identify a conformal field theory associated 
with the nonperturbative lump, and thereby independently verify the 
fermionic nature of its bound states using worldsheet techniques. 

Fermions in the second category are constructed out of gauge fields 
obtained from compactification. These excitations propagate in all 
the bulk noncompact directions. For example, the bosonic string
compactified on the $SO(32)$ group lattice\foot{The authors of \CasherRA\
(see also \EnglertBU\ and references therein) have also studied the bosonic
string compactified down to 10 dimensions group lattices with closely
related motivations.} has a   
potentially rich spectrum of fermions. 

In closing we  note that  Bergman and Gaberdiel \refs{ \BergmanKZ,
\BergmanKM} and Klebanov and Tseytlin \refs{  \KlebanovYY, \KlebanovCH}
have pointed out several years ago 
that strings stretching between $|Dp, +\rangle$ and $|Dp, -\rangle$ branes, 
in type 0A and 0B theory, are fermionic. Such states appear to be fermions 
made out of bosons in a manner distinct from the constructions 
studied in this paper. It would be interesting to understand this better.

\newsec{Spin from Isospin}

In this section we review the Jackiw-Rebbi-Hasenfratz-'t Hooft
construction of fermionic excitations in a 3+1 dimensional bosonic
gauge theory.

Consider an $SU(2)$ gauge theory in 3+1 dimensions  with a single 
real adjoint field $\ph$. The action for this system is  
\eqn\symlag{
S= \frac{1}{g_{\rm{YM}}^2}  \int d^4x \hbox{Tr} \left(
-\frac{1}{2} F_{\mu\nu} F^{\mu\nu}  - D^\mu \ph D_\mu \ph
 \right)
}
\symlag\ may be spontaneously broken to $U(1)$ by a constant $\ph$
vacuum expectation value, 
let $(\ph^1)^2+ (\ph^2)^2 + (\ph^3)^2=v^2$ in the vacuum. 
The 't Hooft-Polyakov monopole that asymptotes to this 
vacuum is 
\eqn\mono{\eqalign{
A_0^a &= 0 \cr
A_i^a &= \frac{\epsilon^{a}_{ij} x^{j}  W(vr)}{r^2} \cr
\ph^a &= \frac{x^{a} H(vr)}{r^2} \cr}}
where $i=1 \ldots 3$ is a spatial vector index, $a=1 \ldots 
3$ is an $SU(2)$ 
isospin vector index and the radial functions $W(vr)$ and
$H(vr)$ are given by
\eqn\func{\eqalign{
W(vr) = 1-\frac{vr}{\sinh vr} \cr
H(vr) = vr\coth vr -1
}}

Note that \mono\ does not transform covariantly under either 
spatial rotations or global isospin rotations. However it does transform
covariantly under a diagonal combination of the two. As we review below, 
this observation allows one to argue that the bound state of the 
monopole \mono\ with a particle  transforming in a 
half integral representation of the gauge group $SU(2)$ is fermionic. 

Consider \HasenfratzGR\ a spinless nonrelativistic 
 particle P, transforming in the $R^{th}$ representation of $SU(2)$, 
interacting with the monopole \mono. 
In an approximation that 
ignores all propagating modes of the fields $A_\m$ and 
$\ph$, the system is described by a Schr\"{o}dinger equation for P in
the background \mono. As pointed out above, \mono\  transforms
covariantly under simultaneous spatial and isospin rotations. 
Consequently the Hamiltonian for the motion of P in the background 
\mono\ commutes with the $SU(2)$ generators
\eqn\angmom{J^a=  \ep^{abc} x^b p^c+ T^a_R}
where $x^b$ is the position operator, $p^c$ its canonical conjugate, 
and $T_R^a$ the isospin generator in the $R^{th}$ representation.
Consequently eigenstates of the Hamiltonian appear in multiplets
of the $SU(2)$ generated by $J$. As states of the 
 entire system, consisting of the monopole plus the particle, are 
independently expected to appear in $SU(2)$ multiplets 
of total angular momentum, Hasenfratz and 't Hooft \HasenfratzGR\ 
proposed that  $J^a$ be identified
with the net angular momentum of the system. As evidence for this 
proposal, Hasenfratz and 't Hooft have demonstrated that, classically, 
the angular momentum  stored in the Yang Mills field when 
a charged particle interacts with a Yang Mills monople 
is indeed given by \angmom\ (see Appendix B). 
According to this proposal, 
the quantum numbers of \angmom\ 
should be interpreted as spin quantum numbers for the monopole - 
charged particle bound state. Consequently, the bound state of a 
monopole with a particle in a half integral representation of the 
gauge group has half integral spin. By the spin statistics theorem, 
such a bound state must be fermionic.  

The angular momentum and fermionic nature of this 
charge-monopole bound state may also be
understood from the long distance $U(1)$ 
description of this system, as we
review in Appendix A.

\newsec{Fermionic 3-String Junctions}

As a concrete application of the mechanism reviewed in the previous 
section we will now argue that the 3-string junction 
described in the introduction, and sketched in Fig. 1 
is a fermion on the world volume of the D3-branes that host it. 
The arguments of subsection 3.1 apply to the 
type 0B theory, type IIB string theory and perhaps even the bosonic
string. In subsections 3.2-3.4 we will separately examine 
the construction in each of these theories. 

\subsec{The 3-String Junction as a Charge-Monopole Bound State}

Consider a D-string stretched between two parallel D3-branes A and B.   
\fig{A D-string stretched between D3-branes A and B. The 3rd brane
M is a spectator.}{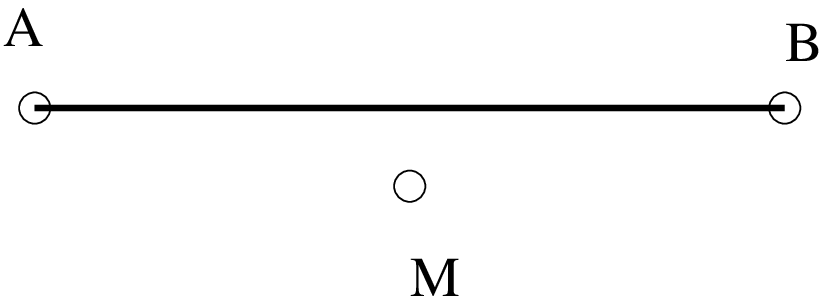}{4.0truein}
When the separation between the D3-branes is small in string units 
the configuration of Fig. 2 is well described as a
monopole \DiaconescuRK\
in the $SU(2)$ theory on the world volumes of branes A and B. Consider 
a string stretching between the branes A and  B and the brane M. Such 
a mode transforms in the fundamental of $SU(2)$. According to the 
arguments of the previous section, a bound state of this  mode 
with the monopole is a fermion. 

It is easy to argue for the existence of at least one such bound
state at weak coupling \WeinbergZT. From the viewpoint of string theory, 
the existence of a bound state is a consequence of the fact that, 
at weak coupling, flux dissolved inside a D-string has less 
energy than an F-string as $\sqrt{({1 \over 2 \pi \apm})^2+
({1 \over 2 \pi  \apm g} )^2} < {1 \over 2 \pi  \apm g}+ 
{1 \over 2 \pi \apm}$.
This bound state is 
the 3-string junction of Fig 1. 

In the general case the 3-string junction between a $(p, 0)$, $(r, q)$
and $(p+r, q)$ is  bosonic if $pq$ is even and fermionic if
$pq$ is odd. This prediction can, in principle, 
  be checked  by a direct computation
of the angular momentum stored in the field of the corresponding 
solution in Yang Mills theory.\foot{ Several authors have investigated
solutions of Yang Mills theory corresponding to quater BPS states. 
\refs{\HashimotoNJ, \HashimotoZS} present a rather explicit construction of 
some 3-string junction solutions for special 
values of $p$, $ q$ and $r$ \HashimotoQH. However, for each of these special 
values,  $pq$ is even: the corresponding solutions are radially symmetric and 
so have zero angular momentum. General 3-string junctions are formally constructed
in \LeeNV, where it is pointed out that the generic 3-string junctions is not 
radially symmetric. In fact, the 3 string junction should be
thought of as a multi centred solution \refs{\ArgyresPV, \KolTW, \GauntlettXZ}. }

\subsec{0B Theory}

Type 0B theory has two varieties of stable D-branes of each dimensionality; the two 
flavors of D-branes are called $|Dp, +\rangle$ and $|Dp, -\rangle$ 
branes, and are 
distinguished by a sign in the fermionic part of the equation that 
defines the corresponding boundary state \BergmanRF \foot{For a recent
review on type 0A and 0B theories, their D-branes see \ThompsonRW.  
We will use the notation of that paper.}.  We will use the following 
important property several times in this paper:
the open strings running between branes or antibranes of the same sign (both
$+$ or both $-$) are always in the Neveu-Schwarz sector, 
whereas strings stretching
between branes or antibranes of different signs are always 
in the Ramond 
sector \BergmanKM.

Let A and B in Fig. 2. both represent $|D3, +\rangle$ branes occupying
spatial dimensions $1, 2, 3$. If follows from the 3-brane world volume
coupling \BilloBG\ proportional to $\int C^+_{0i} B_i$, where $B_i$ is the worldvolume 
magnetic field,  that lines of magnetic flux on a $|D3 + \rangle$ brane
carry $|D1 +\rangle$ charge. 
Consequently, the $SU(2)$ monopole on these 3-branes represents a  
a $|D1, +\rangle$ brane stretching between A and B, in, say the 4 direction. 
If M is also a $|D3, +\rangle$ brane, as in section 4.1, 
the 3 string junction is a stable fermionic bound state of purely bosonic 
degrees of freedom propagating on the $|D3, +\rangle$ branes.  

The fermionic nature of this 3-string junction may also be understood from the 
point of view of the string worldsheet. 
According to the rules stated above, the 
fundamental string from M to the $|D1, +\rangle$ brane is in the
Neveu-Schwarz sector.
Neveu-Schwarz sector fermions in Neumann-Dirichlet directions (in this case $\psi^1,
\psi^2, \psi^3$, $\psi^4$) are integrally moded; in particular the
mode expansion for each of these fermions includes a zero mode. However the
$x^4$ coordinate at the ends of the D-string (on A and B) are fixed and it 
is plausible these boundary conditions project out the  $\psi^4$ zero mode, 
but retain the other three. In which case all open string states appear 
in representations generated by the  $\psi^1,\psi^2, \psi^3$ zero mode
algebra, i.e. are $SO(3)$ spinors.
 
On the other hand, if M is a $|D3, -\rangle$ brane, strings from M to A or
B are fermionic \BergmanKM. The 3 string junction of Fig. 1 is then 
a bound state of the monopole with a fermion in the fundamental of 
$SU(2)$,
and is consequently a boson! From a worldsheet viewpoint this 
is simply a consequence of the fact that Ramond 
sector fermions  in  Neumann-Dirichlet directions, 
(in particular $\psi^1,\psi^2, \psi^3$), are half integrally moded, and so
do not possess a zero mode.

\subsec{IIB Theory}

3-string junctions in IIB theory appear in ${1 \over 4}^{\rm{th}}$ BPS 
multiplets of the $\CN=4$ super algebra on the world volume  of the 3-branes. 
The assignment of half integral spins to the purely bosonic states in this multiplet 
is consistent with this super multiplet structure, as we explain 
in Appendix C using the analysis of \LeeNV\ (see also \BergmanGS).

\subsec{The Bosonic String }

The situation is much more confusing in the Bosonic String. At least 
at leading
order in $\apm$, the open string tachyon couples to gauge and scalar fields 
only quadratically, and so may consistently be set to zero. Consequently, 
when all brane separations are small in string units, the $SU(2)$ 
monopole with the tachyon set to zero is an approximate solution of the 
equations of motion on the D3 branes. The bound state of this monopole with
a fundamental quantum, described in section 3.1,  is a fermionic excitation
in  bosonic string theory.

However, the interpretation of the monopole as a stretched D-string, 
and correspondingly the above configuration as a 3-string junction is
confusing, and may be incorrect\foot{We would like
to thank the participants of the Amsterdam workshop for emphasizing this to
us.} in the bosonic string. 
The energy of the monopole matches the expected energy of a stretched
D1-brane. Further the low energy dyonic excitations of the monopole match the flux
excitations  expected of stretched D1-branes. However the monopole is 
charged, and in particular appears in two varieties, whereas bosonic D-strings are
unoriented and uncharged. 

It is possible that the end points of D-strings on D3-branes are charged, even though 
the the D-strings themselves are uncharged. That certainly seems to be 
true of non BPS branes in type IIA theory\foot{We thank 
Ashoke Sen for explaining this to us.}. 
Consider a D-string and a coincident anti D string ending on a 3-brane-
anti 3-brane system in IIB theory. The D-string could end on either the
3-brane or the anti 3-brane, and the same is true of the anti D-string. 
Of these four possible boundary conditions, the two configurations 
in which both D-strings end on the same 3-brane are not invariant under $(-1)^{F_L}$ 
under which D-branes and anti D-branes are interchanged. The remaining
two boundary conditions are invariant under $(-1)^{F_L}$, and so can be 
modded out by $(-1)^{F_L}$. Consider the 
configuration in which the D-string ends, with a positive magnetic charge 
on the 3-brane, and the anti D-string ends, also with a positive magnetic
charge\foot{The Chern Simons couplings on antibranes have a relative
minus sign compared to those on D-branes. Consequently an anti D-string
ends on an anti 3-brane with the same charge as a D-string on a 3-brane.},  
on the anti 3-brane. Orbifolding this configuration by $(-1)^{F_L}$ 
gives a non BPS D-string ending on the non-BPS 3-brane of IIA theory. 
The end point of this D-string has positive magnetic charge under the
centre of mass $U(1)$ that survives the orbifold projection. 
Orbifolding the other $(-1)^{F_L}$ invariant configuration 
(the D-string ending on the anti 3-brane, and the
anti D-string ending on the 3-brane) leads to a non
BPS D-string ending on a non BPS 3-brane with a negative magnetic charge. 

Thus uncharged and unoriented non-BPS D-strings can end on non BPS 3-branes
in two inequivalent ways. These endpoints are associated  with either
positive or negative magnetic charge in the 3-brane theory.
It would be interesting to understand if this was true even in the bosonic string theory.

Another confusing question surrounding this configuration is the following: 
what happens to this fermion, after the D3-branes that host it have
disappeared due to tachyon condensation? This question has a perhaps
sharper analogue in type 0A theory. Consider an unstable 
3-brane of the $+$ variety, $|\hat{D3}, +\rangle$, 
joined to an unstable 3-brane of the $-$ variety, $|\hat{D3}, -\rangle$ 
by a single fundamental string. What happens to this fermionic excitation after the two 
unstable 3 branes have decayed due to tachyon condensation? In both 
cases the resulting configuration would appear to be an infinitely long  
open string with fermionic nature. It would be interesting to understand this
better.  

\newsec{Spin from Isospin in higher dimensions}

As reviewed in section 2, a bound state of 
an $SU(2)$ monopole and a fundamental quantum in 3+1 dimensions should 
be regarded as a fermion. In four or higher dimensions finite energy 
solitons analogous to the 't Hooft-Polyakov monopole do not 
exist\foot{This may 
be understood as follows. Consider a scalar field (charged under the 
gauge group) that winds at infinity. The angular derivatives of this 
field decay at infinity like ${1 \over r}$ ; in order that all 
covariant derivatives decay sufficiently rapidly, the gauge field 
is forced to take a specific value at $\CO({1\over r})$, implying 
$F \sim \CO( {1 \over r^2})$, and a configuration energy  
$E\approx \int d^d x F^2 \sim \int {d^dx \over r^4} $, divergent 
for $d \geq 4$, where $d$ is the spatial dimension.}. However 
instantons of the $d$ dimensional gauge theory are
localized finite energy field configurations. These configurations 
are classified by $\pi_{d-1} (G)$, where $G$ is the gauge group
and are stabilized by the winding of the gauge group at infinity.
 
In this section we will study instantons of $U(N)$ and $SO(N)$ gauge
theories, and argue 
that their bound states with quanta transforming in certain 
representations of the gauge group are often fermionic, generalizing
the mechanisms for producing spin from isospin reviewed in section 2. 

While instantons are scale invariant in 4 dimensions, 
energetics drives them to shrink to a point in higher dimensions. 
In the applications we have in mind, we expect these instantons to 
be stabilized at short distances by stringy effects. 
Since all our conclusions will follow from long distance 
topological considerations, the details of this stabilization will not
bother us. 

\subsec{Fermions from $U(N)$ instantons}

The even homotopy groups of $U(N)$ for $N \geq 2$  are trivial
($\pi_{i}( SU(N))=0$ 
for all even $i \geq 2$)  while the odd homotopy groups are  
integer valued  ( $\pi_{i}( SU(N))= Z$ for all odd $i \geq 3$).
Consequently interesting $U(N)$ instantons exist on even dimensional
spaces; the quantized instanton number of a configuration in $d=2m$
dimensions is given by ${1\over m! (2\pi)^m } \int F \wedge F \wedge \ldots 
\wedge F$. 

Note also that $U(N)$ homotopy groups are stable for sufficiently large
$N$. Consequently, for $N$ sufficiently large compared to $d$ a $U(N+1)$
instanton is isomorphic to the trivial embedding of a $U(N)$ instanton 
into $U(N+1)$.

In this subsection we study $U(N)$ instantons in $d$ spatial
dimensions, where $d$ is even.
The stability property of $U(N)$ homotopy groups 
permits us to study any particular convenient large value of $N$ for 
every spatial $d$. Most of our results will then be easy to generalize 
to other large values of $N$.  We find it  convenient to choose
$N=2^{{d-2 \over 2}}$ as 
this allows us to use the  Atiyah Bott Shapiro 
construction \abs\ (see also \WittenCD )  of this instanton at infinity.

Consider an $SU(2^{{d-2 \over 2}})$ theory in $d$ (even)  spatial dimensions. 
Let $\g_i$ ( $i = 1 \ldots d-1$) represent the $2^{{d-2 \over 2}} \times 
2^{{d-2 \over 2}}$  $SO(d-1)$ $\Gamma$ matrices. Define 
$\g_d =-i I$. Consider an instanton in this theory whose gauge 
field takes the form
\eqn\pgai{A_\m= -i \p_\m U U^{-1} f(x^2)}
where
\eqn\uain{U(x)={\g_\m x^\m \over |x|}}
and $f$ tends to unity as its argument becomes large. 
$(\g_\m)_{\a {\dot \a}}$ is an $SO(d)$ invariant\foot{This follows from the
observation that the set of matrices $\G_i \otimes \s_1 , I \otimes \s_2$  constitute a
set of $SO(d)$ $\G$ matrices whose chirality matrix is $ I \otimes \s_3$.}, where  
where $\a$ is a chiral $SO(d)$ spinor index and ${\dot \a}$ is an
antichiral $SO(d)$ index. 
Therefore, under a spatial rotation $R$, 
\eqn\utr{U \r S(R) U {\bar S}^T(R)}
where $S(R)$ and ${\bar S}(R)$ respectively are the chiral 
and antichiral
spinorial matrices corresponding to the rotation $R$. Consequently, 
under the same rotation, 
\eqn\utr{A_\m \r S(R) R_\m^\n  A_\n S^{-1}(R).}
Thus, $A_\m$ transforms covariantly under the rotation $R$
accompanied by a simultaneous large gauge transformation by the 
$U(2^{{d-2\over2}})$ matrix $S(R)$. 
Consequently, as in section 2, the conserved angular 
momentum $J_{\m\n}$ of the system is 
\eqn\cam{J_{\m\n} = x_\m p_\n -x_\n p_\m  +T_{\m\n}}
where $T_{\m\n}$ are the generators of an $SO(d)$ subgroup of 
$SU(2^{{d-2 \over 2}})$. 

As an immediate consequence, the bound state 
 of a fundamental scalar with this instanton is a chiral spinor, and so 
is fermionic. This state pairs up with the corresponding
 anti-instanton bound state, an antichiral spinor, to form a full 
$(d,1)$ dimensional spinor.  

Several remarks are in order.

\item{1.} The $SO(d)$ subgroup of $SU(2^{{d-2 \over 2}})$ that appears 
in \cam, is determined by the following defining property: 
the fundamental  representation of $SU(2^{{d-2 \over 2}})$
transforms as a chiral spinor of $SO(d)$. In general an arbitrary 
representation of $SU(2^{{d-2 \over 2}})$ will decompose into 
several representations of $SO(d)$. This decomposition may be determined 
as follows. The $d \over 2$ Cartan 
generators of $SO(d)$ may
be determined  as functions of the $SU(2^{{d-2 \over 2}})$, as
the explicit form of generators of both groups is available in 
the $2^{{d-2 \over 2}}$ dimensional representation. Consequently the 
$SO(d)$ weights, and representation content of the bound state with 
any $SU(2^{{d-2 \over 2}})$ quantum may be determined.

\item{2.} As \pgai\ is pure gauge at infinity, the instanton does not 
break any gauge symmetry unlike the monopole \mono. At infinity, 
$A_\m$ in \pgai\ may be locally set to zero in coordinate patches.
Consequently, the bound state of an instanton with a particle
in the $R^{th}$ representation of $U(N)$ continues to transform in the
$R^{th}$ representation of the unbroken $U(N)$.

\item{3.} As remarked at the beginning of this subsection, 
 it is straightforward to generalize the considerations of this 
subsection to  $SU(N)$ theories with   
$N > 2^{{d-2 \over 2}}$. The instantons of this subsection may 
simply be embedded into a                block diagonal
$SU(2^{{d-2 \over 2}})$ subgroup of $SU(N)$.

\item{4.} It is instructive to separately 
study the simplest special case, namely the familiar  
$SU(2)$ instantons in $d=4$.
\uain\ takes the form 
\eqn\uai{U(x)={\s_\m x^\m \over |x|}}
where $x^\m$, ($\m = 1 \ldots 4$) are the spatial coordinates,   
$\s_i$ are $i =1 \ldots 3$ are the Pauli matrices,  and $\s_4=-i I$.
The rotational symmetry group of $R^4$ is $SO(4) = SU(2)_L \times 
SU(2)_R$. Repeating the general arguments of this subsection, we find that 
the true angular momentum generators are
\eqn\comgen{K_a'=K_a+T_a, ~~~ L_a, ~~~a = 1 \ldots 3}
where $T_a$ are the $SU(2)$ gauge generators, and $K$ and $L$ are the
generators of  $SU(2)_L$ and $SU(2)_R$ \foot{Specifically, $K$ and $L$ 
are linear combinations of the usual $SO(4)$ generators 
\eqn\dolk{\eqalign{K_a=&{\half \ep_{aij}L_{ij}+L_{0a} \over 2} =
\frac{1}{4}\eta^{a\mu\nu}L_{\mu\nu} =\half \eta^{a\mu\nu} x_\m p_\n \cr
L_a=&{\half \ep_{aij}L_{ij}-L_{0a} \over 2}
= \frac{1}{4} \bar{\eta}^{a\mu\nu}L_{\mu\nu}=\half  \bar{\eta}^{a\mu\nu}
x_\m p_\n
 }}
where $\eta^{a\mu\nu}$ and $\bar{\eta}^{a\mu\nu}$ are the 't Hooft symbols
defined and described in equations A5-A15 of \tHooftFV.}. 
In the classical approximation, it is possible 
to independently compute the angular momentum stored in the Yang Mills 
field for the configuration consisting of a particle in the $R^{th}$ representation
of $SU(2)$  placed in the background of a Yang Mills instanton. We 
present the relevant computation in Appendix B; we find agreement 
with \comgen.

\subsec{$SO(N)$ instantons}

In this subsection we will study bound states of charged quanta 
with $SO(N)$ instantons in $8k+1$ spatial dimensions, and find that several 
of these are fermionic. 

$\pi_{8k}(SO(N))=Z_2$ for $N \geq 8k+2$; consequently $SO(N)$ theories 
in $8k+1$ spatial dimensions possess a unique nontrivial instanton.
Consider, for concreteness, 
 an $SO(16)$ gauge theory in 9+1 dimensions. Recall that 
the 9 $16 \times 16$ $\Gamma$ matrices $\Gamma_\m$ 
of $SO(9)$ may be chosen to be real and symmetric; we make such a 
choice. The unique  instanton may be put into the form\foot{We thank
E. Witten for discussions in this connection.} 
\eqn\uain{A_\m= -i \p_\m O O^{-1}f(x^2), ~~~
O(x)={\G_\m x^\m \over |x|.}}
where $O(x)$ is an $SO(16)$ group element, and $\lim_{x^2 \r \infty} f(x^2)
=1$.  $(\G_\m)_\a^\b$  is an $SO(9)$ invariant where $\m$ is a vector index
and $\a$ and $\b$ are $SO(9)$ spinor indices. Imitating the logic of the 
previous subsections we conclude that the angular momentum 
operator in the instanton background is
\eqn\cam{J_{\m\n}= x_\m p_\n -x_\n p_\m  +T_{\m\n}} 
where $T_{\m\n}$ are the generators of an $SO(9)$ subgroup of 
$SO(16)$ gauge group. $SO(9)$ is embedded into $SO(16)$ 
in the manner so that 
the vector of $SO(16)$ transforms as a spinor of $SO(9)$; as in the 
previous subsection this completely specifies the $SO(9)$ representation
content of any $SO(16)$ multiplet. 
Consequently, for example, a zero orbital angular momentum 
bound state of a vector with this instanton is fermionic. It transforms
as a spinor under angular momentum and as a vector of the 
$SO(16)$ gauge group.

The instanton of this subsection may be 
embedded into the natural or trivial 
$SO(16)$ subgroup of $SO(N)$ for $N \geq 16$, 
to yield the unique $SO(N)$
instanton. All considerations of this subsection 
generalize to this case.

\newsec{Applications in String Theory}

In the previous sections we have presented a construction of 
fermions out of bosons. This construction applies to 
any system that includes the appropriate non Abelian gauge fields in the 
appropriate dimensions. 
In particular, it finds applications in  
perturbatively bosonic string theories (e.g. type 0A theory and the 
bosonic string), permitting the identification of nonperturbative 
fermions in these theories, as we describe in this section. 
In some examples we are able to identify the world sheet conformal field theory 
that corresponds to our construction and so are able to verify the fermionic
nature of these excitations using world sheet techniques. 

Gauge fields in perturbative string theory arise either from open strings 
on D-branes or from closed strings via compactification on group lattices. 
In subsection 5.1 and 5.2 we utilize the gauge fields on D branes in
bosonic string theories to construct fermions propagating on their 
world volume. In subsection 5.3 we utilize gauge fields from
compactification to construct fermions propagating in space time.

In subsection 5.1 we study type 0A theory in the presence of $N+1$ 
$|Dp, +\rangle$ branes. The perturbative spectrum of both open and 
closed strings in this background is purely bosonic. According to the 
construction presented in section 4.1, the bound state of an instanton 
in the world volume theory  of the
$N$  $|Dp, +\rangle$ branes, with a  
fundamental string  stretching from the $N$ branes to the 
last remaining brane, is fermionic. However an instanton on a $|Dp, +\rangle$ brane 
is simply a $|D0, +\rangle$ brane. Using this observation and world sheet
techniques, it is easy  to directly verify the fermionic nature of the
corresponding system.
 
The general arguments of section 4 also predict that the same worldvolume
configuration, namely the bound state of a string and an instanton on D
branes, is fermionic even in the bosonic string
theory (the open string tachyon is approximately unsourced by an instanton
that is large in string units). In the bosonic string, however, there seems 
no reason to associate an instantonic gauge field configuration with a D0
brane, or any other familiar worldsheet theory. Consequently we do not have
an independent verification of the fermionic nature of this state.

In subsection 5.2 we consider type O theory; an orientifold 
of Type 0B theory that 
includes 32 D9 and 32 anti D9 branes in its background. 
We consider the $SO(32)$ instanton 
constructed out of the gauge fields on
the D9 branes. It is natural to conjecture, in analogy with Type I theory, 
that this instanton represents the stable 0-brane of type O theory. 
According to the arguments of the previous section, the 
bound state of such an instanton with a bi-fundamental open string 
(a string stretching between a brane and an anti-brane) is fermionic. 
Using the 
conformal field theory description of the stable  0-brane, 
this prediction is easily verified. 

In subsection 5.3 we point out that the constructions of section 4 imply
that bosonic string theories compactified on group lattices have a 
potentially rich spectrum of fermions. We spell this out in a particularly
interesting example; the bosonic string compactified on the $SO(32)$ group 
lattice.

\subsec{The D0-Dp system in Type OA theory }

In this subsection we study a string theoretic implementation of the 
construction of fermions from $SU(N)$ instantons presented in subsection
4.1. Using worldsheet techniques we find independent confirmation of the 
fermionic character of our state. 

Consider $2^{{p-2 \over 2}}+1$ $|Dp, +\rangle$  parallel branes, 
aligned along directions $x^1, x^2,\ldots x^p$, in type 0A theory. 
Let the last of these branes be slightly separated, in transverse space,  from the other  
$2^{{p-2 \over 2}}$ branes, all of which are coincident. 
Consider an $SU(2^{{p-2 \over 2}})$ bundle with a single unit instanton
charge on the world volume of the coincident branes. Such an instanton carries 
the charge of a single $|D0, +\rangle$ brane \BilloBG, and so represents
a $|D0, +\rangle$ brane dissolved in the $2^{{p-2 \over 2}}$ coincident 
$|Dp, +\rangle$ branes. Consequently, the quantum numbers of a string
stretching from this instanton to the isolated $|Dp, +\rangle$ must be
identical to those of a string running from a $|D0, +\rangle$ to the 
isolated $|Dp, +\rangle$ brane\foot{Some aspects of the 
p-p+4 system of type 0B theory were studyed in \CostaQX.}. 

It is easy to formally\foot{The rest of this subsection is formal as 
charge conservation forbids a single string from ending on a 0-brane,
unless the later is dissolved in a higher dimensional brane, in which case
the fundamental string charge is carried by lines of electric flux which
spread out to infinity.} verify that a $|D0, +\rangle$ brane sitting on top
of a $|Dp, +\rangle$, and connected to it with a fundamental string, is 
a spinor on the $p$ spatial dimensional world volume of the brane.
As we have remarked above (see subsection 3.2) strings between 
$|D0, +\rangle$ brane and the isolated $|Dp, +\rangle$
are always in the Neveu-Schwarz sector. 
World sheet Neveu-Schwarz fermions in directions with  Neumann-Dirichlet
boundary conditions,   $\psi^1, \psi^2,
\ldots \psi^p$ in this example, are always integer moded. Consequently 
0-p  open string states appear in degenerate multiplets obtained by
quantizing the zero modes of the fermions $\psi^1, \psi^2,
\ldots \psi^p$, and so are  $SO(p)$ spinors. As $0-p$
strings are oriented, these spinors are complex. The GSO projection, which
 retains only state of even worldsheet fermion number, projects out the 
odd chirality component states of this spinor, retaining a chiral   
$SO(p)$ spinor, in precise agreement with the analysis of section 4.1.

\subsec{The D-particle in Type O theory}

In this subsection we study a string theoretic implementation of the 
construction of fermions from $SO(N)$ instantons presented in subsection
4.2. We verify the fermionic nature of our state using worldsheet
techniques. 

Recall that the stable D-particle of type I theory,  discovered by Sen
\SenTT,  has the `charge' of an $SO(N)$ instanton 
residing on the background D9-branes of Type I theory \WittenCD. 
type O theory possesses two stable D particles, $|\hat{D0}, +\rangle$ and 
$|\hat{D0}, -\rangle$, obtained by taking the orientifold projection of 
the two unstable D particles of 0B theory. As in \SenTT, the action of the 
orientifold projects out the 0-0 tachyon, rendering these particles
stable. It is natural to conjecture that these stable D-particles
carry the `charge' of $SO(N)$ instantons on $|D9, +\rangle$ and 
 $|D9, -\rangle$ branes respectively. 

Consider type O theory with 32 $|D9, +\rangle$ branes and 32 anti $|D9, +\rangle$ branes
in its background. The perturbative spectrum of both open and closed strings in this
background is purely bosonic. According to the construction of section
4.2, an $SO(32)$
instanton on the $|D9, +\rangle$ branes, bound to a bifundamental 9 -
${\bar 9}$ string, is fermionic. As in the previous subsection (and
according to the conjecture of the previous paragraph) the quantum numbers
of such a bound state must be identical to those of the formal state
obtained by quantizing a single $|\hat{D0}, +\rangle$ - anti $|D9, +\rangle$
string. That is easy to verify. The relevant $(0,9)$ strings 
are always in the Neveu-Schwarz sector and have no GSO projection.
The world sheet fermions $\psi^i$, $i= 1, \ldots 9$ have
Neumann-Dirichlet boundary conditions, and are consequently integer moded. 
Thus 0-9 string states appear in multiplets obtained by quantizing the zero 
modes of $\psi^i$, $i= 1, \ldots 9$, and so are $SO(9)$ spinors. Since these
strings are unoriented the corresponding spinors are real.

The fermions $\psi^i$ also carry an anti 9-brane Chan-Paton index, and 
so are vectors\foot{All 0-9
string states in the Neveu-Schwarz sector are massive (this follows from the zero point
energy of these strings). Consequently, states appear in multiplets
obtained by quantizing the 
zero modes of any one of the 32 possible strings, and so transform in the
vector of $SO(32)$. Strings in the Ramond sector are massless in their ground
state. Consequently states in the Ramond  sector appear 
 in multiplets obtained by quantizing the zero modes of $all$ 32
strings. In the Ramond sector only $\psi^0$ is integrally moded. The
spinorial character of the
type I D particle follows by quantizing $\psi^0$ for all 32 0-9 strings.
We thank S. Mukhi for discussions in this connection.} 
under $SO(32)$, in precise agreement with the analysis of section 4.2.

\subsec{Fermions in the Bosonic String on the $SO(32)$ group lattice}

As we have remarked above, the construction of section 4 
implies the existence of a spectrum of spacetime fermions  
in bosonic string theories compactified on $U(N)$ or $SO(N)$ group
lattices. In this subsection we consider a special case; the bosonic
string compactified down to 10 dimensions on the $SO(32)$ group lattice. 
This particular compactification of the bosonic string seems especially 
interesting to us. Firstly, the left moving sector 
of this bosonic string is identical to the left moving sector of 
the $SO(32)$ Heterotic string. Secondly, in a speculative but  intriguing conjecture, 
Bergman and Gaberdiel have suggested that this compactification of the
bosonic string  is S-dual to type O theory \BergmanRF.

Apart from the metric, $B_{\m\n}$ the dilaton, and the uncharged tachyon, 
the low lying spectrum of this compactification 
includes $SO(32) \times SO(32)$ gauge bosons, a set of tachyons in 
the bi-fundamental of $SO(32)\times SO(32)$, and the lattice deformation
modes : massless scalars in the bi-adjoint representation of $SO(32)\times
SO(32)$. The massive spectrum includes Hagedorn towers of states 
with each of these $SO(32) \times SO(32)$ quantum numbers.

Consider a nonperturbative field configuration in this theory that is 
characterized by the charge of a $Z_2$ instanton in one of the $SO(32)$
gauge groups, as studied in section 4.2. According to section 4.2, the
bound state of such an `instanton' with, for instance, the bifundamental 
tachyons or massive string states,  is fermionic. 
Consequently, this compactification of the bosonic string has a potentially
rich spectrum of fermions. 

Unfortunately we do not have a detailed understanding of any of the states
of this subsection. However, one could speculate 
that tachyon condensation
in this system has an endpoint, and that the nonperturbative
fermions described in section become lighter as the tachyon condenses.

\newsec{Discussion}

We have generalized the Jackiw-Rebbi-Hasenfratz-'t Hooft construction of
fermions from bosons, and argued that this construction may be used to 
identify a rich nonperturbative spectrum of fermions in
perturbatively bosonic string theories. In some examples 
we are able to find an alternative worldsheet description of these states,
and independently confirm their fermionic character.

Perhaps the observations in this paper make it a little more plausible 
that the bosonic string is connected to the M theory mainland. 
In this connection we find the bosonic
string compactified on the $SO(32)$ lattice especially intriguing. As argued
in section 5.3, this theory has a rich spectrum of 10 dimensional
fermions. Further, its worldsheet theory is closely related to that of the 
$SO(32)$ Heterotic string. 
Indeed, it is possible to act on the conjectured 
relationship between 0A and IIA theories \refs{\BergmanKM, \CostaNW, \GutperleMB}
with a  sequence of very speculative dualities, to arrive at a
conjectured relationship between the bosonic string theory on 
the $SO(32)$ lattice and the $SO(32)$ heterotic string. Wild as these ideas sound, we
believe they are worth exploring.   

Finally, at a more mundane level, it would also be interesting to address some of
the unanswered  technical questions touched upon in this paper. What are the rules
that determine when  D-branes are allowed to end on other D-branes in the
bosonic string theory? What charges, if any, do these end points carry? What 
is the fate of fermions hosted on unstable D-branes after the branes have
decayed into nothing?  Is it possible to find a simple conformal
field theory that describes a finite D-string stretching between two 3
branes?

\centerline{\bf Acknowledgements} We are grateful to 
D. Bak, O. Bergman, A. Dabholkar, 
R. Gopakumar, G. Mandal, J. Polchinski, T. Ramdas, S. Mukhi, 
F. Schaposnik, E. Silverstein,  A. Strominger, 
D. Thompson, S. Trivedi, S. Wadia, 
E. Witten, and the participants of the 
Amsterdam Summer Workshop for useful discussions. We are 
especially grateful to Ashoke Sen for several extremely useful 
discussions and suggestions. S.M. would like to thank ITP Santa Barbara, 
TIFR Mumbai and the Amsterdam Summer Workshop for hospitality during 
various stages of completion of this paper.  This work was supported
in part by DOE grant DE-FG02-91ER40654, a Harvard Junior Fellowship
and Fundacion Antorchas. The work of J.R.D is supported  by NSF grant PHY97-2202.

\appendix{A}{Charge-Monopole Angular Momentum in the effective Electrodynamics}

Consider an $SU(2)$ gauge theory \symlag\ in the presence of a 
't Hooft Polyakov monopole with asymptotic Higgs vev
$v$. Outside the monopole core ($r \gg {1 \over v}$) and 
for processes of energy scale $\omega \ll v$
\symlag\ is effectively 
a $U(1)$ gauge theory. Consequently, as $v$ is taken to infinity
a $U(1)$ description of the monopole \mono\ is  arbitrarily good 
everywhere. Indeed, in a `unitary gauge', 
in which the $\ph$ field is everywhere 
aligned along the $3$ direction,  
\mono\ and \func\ reduce
\eqn\monon{\eqalign{
A_0^a &= 0 \cr
A_i^a &=\d^{3a}{(1-\cos(\t)) \over r} {\hat  \ph} \cr
\ph^a &= v \d^{a3},  \cr}}
the potential of a Dirac monopole of charge 
$g=\int_{S^2} {\vec B}^3.{\vec ds}= 4 \pi$ (where $\vec B$ is the
magnetic field) with a Dirac string along the $-{\hat z}$ axis.  

An $SU(2)$ doublet consists of two particles of unbroken $U(1)$ charge  
$\pm \half$. As $\half  \times 4 \pi= 2 \pi \times 1$, 
these particles saturate the Dirac quantization condition with the 
monopole \monon.  Now, according to Saha's classic calculation \Saha, 
a monopole of charge $g$  and an electron of charge
$e$ separated in space by the position vector ${\vec r}$ have angular
momentum
\eqn\sam{{\vec J}= {eg \over 4 \pi} {\hat r}.}
Consequently  the monopole \monon\ separated from either of the two particles
from the $SU(2)$ doublet by a position vector ${\vec r}$ 
has the half integral
angular momentum ${\vec J}={\hat r}$, in agreement with the analysis of 
subsection 2.1.

We have demonstrated that the angular momentum  of a monopole-fundamental
bound state is half integral.  An appeal to the spin- statistics theorem  
then demonstrates the fermionic character of these bound states. In fact,  
within an effective $U(1)$ theory, 
Goldhaber has also given a remarkably 
simple direct argument for the fermionic statistics of such 
bound states \GoldhaberDP.

\appendix{B}{The Angular Momentum of Bound States }

\subsec{Angular Momentum of the Charge-Monopole system in 3+1 dimensions}

As reviewed in Sec 2.1, Hasenfratz and 't Hooft have proposed that the 
angular momentum of a particle in the $R^{th}$ representation of 
$SU(2)$ in the presence of an $SU(2)$ monopole is given by 
\angmom. As evidence for this proposal, Hasenfratz and 't Hooft have  
computed the angular momentum
\eqn\tam{L_{total}^a=L_{particle}^a+L_{field}^a
=\ep^{abc}x^b (m {\dot x}^c)+ 2 \int d^3x  \ep^{abc}
\hbox{Tr}\left( F_{0i} F_{ib}x^c +D_0 \ph x^c D_b \ph \right) }
of the configuration under consideration, in the classical limit.
In their computation, 
the magnetic field $F_{ib}$ and $D_b \ph$ in \tam\ take their background 
values \mono, while the electric field $F_{0i}$ and $D_0 \ph$ 
are sourced by the charged particle, and are determined by the equations 
of motion. 
Thus \tam\ may be evaluated as a function of the particle's
position in space $x$, and orientation in group space $T^a$, 
yielding \HasenfratzGR\ ${L_{field}^a=
T^a+e \ep^{abc} x^b A_c^k T^k.}$
Consequently, 
\eqn\tam{L_{total}^a=L_{particle}^a+L_{field}^a
= T^a+\ep^{abc}x^b p^c}
where $p^a=m {\dot x}^a+e A_a^b T^b$ is the momentum conjugate to 
$x^a$, confirming the identification of $J^a$ with the 
total angular momentum of the system.

In subsection B.1 below we mimic the procedure outlined above to 
compute the angular momentum of a charged particle in the presence of a 
Yang Mills instanton in 4+1 dimensions. We find that the answer is indeed 
given by \comgen, as predicted in subsection 4.1. We regard this
computation as a check on the arguments of section 4. 

\subsec{Angular Momentum of the Charge Instanton system in 4+1 dimensions}

Consider the pure $SU(2)$ gauge theory in 4+1 dimensions 
\eqn\pymlag{
S=- {1 \over 4 g_{YM}^2} \int d^5 x F^a_{AB}F^{aAB}.
}
where $A, B$ are spacetime indices that run  from $0, 1,\ldots 4$, 
and $a, b, \ldots$ are gauge indices that run  from $1, 2, 3$.
The 4 dimensional self dual instanton is a static solution of \pymlag\
and is given by
\eqn\inst{\eqalign{
A_\mu^a &= 2 \frac{\eta^{a\mu \nu} x^\nu}{x^2 + \lambda^2}, \cr
        &= 2 \eta^{a\mu\nu} x^\nu W \cr
F_{\mu\nu}^a &= 4\eta^{a\mu\nu} W(x^2 W  -1).
}}
where $\mu, \nu, \ldots$ are spatial indices and take values from $1, \ldots 4$.
Here $W(x) = \frac{1}{x^2 +\lambda^2}$, $x^2 = x^\mu x_\mu$ and
$\lambda$ is the scale of the instanton.
$\eta^{a\mu\nu}$ is the 't Hooft symbol. 

Consider a classical test particle with isospin charge vector $T^a$ at the point
$y^\mu$. This creates a field $A_0^a$ according to the equation
\eqn\eqmotion{
D^{ab}_\mu F_{\mu 0}^b = g_{YM}^2 T^a \delta^4(x-y).
}
(recall that $g^2_{YM}$ has dimensions length) which in turn gives rise to an
angular momentum contribution given by
\eqn\angm{
L_{\alpha\beta} = \frac{1}{g_{YM}^2}
\int d^4x ( x_\alpha F^a_{0\rho} F^{a}_{\rho\beta}
- x_\beta F^a_{0\rho} F^{a}_{\rho\alpha} ).
}
Substituting the instanton configuration from \inst, and performing
simplifications which make use of the identities listed in Eqns
A5-A15 of \tHooftFV\ we obtain
\eqn\angma{\eqalign{
L_{\alpha\beta} = \int &d^4x \left[
(\eta^{a\beta\rho}\partial_\rho A_0^a x^\alpha
-\eta^{a\alpha\rho}\partial_\rho A_0^a x^\beta) 4 W(x^2W -1) \right. \cr
&- \left. ( \eta^{a\beta\rho}x^\rho x_\alpha -
\eta^{a\alpha\rho}x^\rho x_\beta ) 16 W^2(x^2W -1)\right].
}}
A covariant way to split these $SO(4)$ generators into $SU(2)_R$ and
$SU(2)_L$ is to project out the self-dual and anti-self dual parts of
$L_{\alpha\beta}$ in \angma. The self dual part is given by
\eqn\angself{\eqalign{
J_{\alpha\beta} &= 
L_{\alpha\beta} + \frac{1}{2}\epsilon^{\alpha\beta\mu\nu}L_{\mu\nu},
\cr
= \frac{1}{g_{YM}^2} &\int d^4 x
\left\{\left[
\eta^{a\beta\mu}( x_\alpha \partial_\mu A_0^a - x_\mu \partial_\alpha
A_0^a) - 
\eta^{a\alpha\mu}( x_\beta \partial_\mu A_0^a - x_\mu \partial_\beta
A_0^a) \right.\right. \cr
&- \left.\left.  \eta^{a\alpha\beta} x^\mu\partial_\mu A_0^a 
\right] 4 W(x^2W -1) 
+ x^2\eta^{a\alpha\beta} 16 A_0^a W^2(x^2W-1)\right\}.
}}
The anti-self dual part is given by
\eqn\anganti{\eqalign{
I_{\alpha\beta} &=
L_{\alpha\beta} - \frac{1}{2}\epsilon^{\alpha\beta\mu\nu}L_{\mu\nu},
\cr
= \frac{1}{g_{YM}^2}\int &d^4 x
\left\{\left[
\eta^{a\beta\mu}( x_\alpha \partial_\mu A_0^a + x_\mu \partial_\alpha
A_0^a) - 
\eta^{a\alpha\mu}( x_\beta \partial_\mu A_0^a + x_\mu \partial_\beta
A_0^a) \right. \right. \cr
&+ \left.  \eta^{a\alpha\beta} x^\mu\partial_\mu A_0^a 
\right] 4 W(x^2W -1)  \cr
&- \left. ( 2\eta^{a\beta\mu}x_\mu x_\alpha - 2 \eta^{a\alpha\mu} x_\mu
x_\beta + x^2 \eta^{a\alpha\beta} ) 16 A_0^a W^2(x^2W-1)\right\}.
}}
The corresponding generators of $SU(2)_L$ are given by 
\eqn\gensul{K'_a={1 \over 4} \eta^{a \m\n}L_{\m\n}={1 \over 8} \eta^{a
\m\n}J_{\m\n}}
and those of $SU(2)_R$ are given by 
\eqn\gensur{L_a={1 \over 4}{\bar  \eta}^{a \m\n}L_{\m\n}={1 \over 8} {\bar
\eta}^{a\m\n}I_{\m\n}.}

We wish to evaluate $J_{\a\b}$ in \angself. Note that the first two terms
in \angself\ can be written as a  total divergence, and vanish upon
integrating. Consequently 
\eqn\angselfa{
J_{\alpha\beta} ={1 \over g^2_{YM}} \int d^4 x \left\{
-4 \eta^{a\alpha\beta} x^\mu\partial_\mu A_0^a W(x^2W -1) 
+ x^2\eta^{a\alpha\beta} 16 A_0^a W^2(x^2W-1)\right\}.}
The equation of motion for $A_0$, \eqmotion,  may be simplified to 
\eqn\eqmotiona{
\partial_\mu\partial_\mu A_0^a + 4 \epsilon^{abc}\eta^{b\mu\nu} x^\nu
W\partial_\mu A_0^c - 8 x^2 A_0^a W^2 = g_{YM}^2 T^a \delta^4(x-y).
}
From \eqmotiona\ we solve for $A_0^a$ as a function of $T^a$ and
derivatives of $A_0^a$, and substitute this solution into the second term
of \angselfa. Upon integrating by parts, all the terms in \angselfa\
involving $A_0^a$ cancel, and we obtain the simple expression
\eqn\fangslef{
J_{\alpha\beta} = (-2T^a y^2 W  + 2 T^a) \eta^{a\alpha\beta}.
}
The $SU(2)_L$ generators are now given by
\eqn\lefang{\eqalign{
K'_a &= \frac{1}{8}\eta^{a\alpha\beta}J_{\alpha\beta}, \cr
&= -T^a y^2 W + T^a \cr
&= \half \eta^{a\m\n}x_\m A_\n^a T^a +T^a.}}
In order to compute the total angular momentum, we add to $K'_a$ the
contribution due to the orbital angular momentum of the particle
\eqn\orbang{K_a^{orb}={1 \over 4}\eta^{a\m\n}(x_\m {\dot x}_\n-x_\n {\dot
x}_\m)=\half \eta^{a\m\n} x_\m
{\dot x}_\n}
to obtain
\eqn\toangself{
K^{\prime total}_a = \frac{1}{2}\eta^{a\mu\nu}x_\mu p_\nu + T^a.
} 
where $p_\mu = m\dot{x}_\mu + A_\mu ^a T^a$ is the canonical momentum
conjugate to $x^\m$. Clearly \toangself\ agrees with \comgen, and 
 demonstrates the mixing of the $SU(2)_L$ part of angular momentum with the 
$SU(2)$ generators of the gauge group.

We now turn to the computation of $I_{\a\b}$,  the anti-self dual part of the
conseved angular momentum. As above we use \eqmotiona\ to  solve for
$A_0^a$ in terms of $T^a$ and its derivatives, and plug that solution into 
the last term in \anganti. After some tedious but
straight forward integration by parts we find that all  terms in \anganti\ involving
$A_0$ cancel, and 
\eqn\fanganti{
I_{\alpha\beta} = 2(2\eta^{a\beta\mu} y_\mu x_\alpha - 2
\eta^{a\alpha\mu}x_\mu x_\beta + x^2\eta^{a\alpha\beta} ) T^a W.
}
It is easy to see that $I_{\alpha\beta}$ is the anti-self dual part of
$(y_\alpha A^a_\beta - y_\beta A^a_\alpha) T^a$ where $A^a$ is the
instanton background. Thus the anti-self dual part of the 
total angular momentum including the anti-self dual part of the
orbital angular momentum is given by
\eqn\toanganti{
L_a = \frac{1}{2} \bar{\eta}^{a\mu\nu}x_\mu p_\nu,
}
in agreement with \comgen.
Note that the anti-self dual part of the conserved angular momentum
does not mix with the generators of the gauge group.

\appendix{C}{Supermultiplet structure 
of 3-string junctions in IIB theory}

In Type IIB theory the 3 string junction of Fig. 1. is expected to appear in a 
$({1\over 4})^{\rm th}$ BPS multiplet of $\CN=4, d=4$ supersymmetry. In
this appendix we will construct this 3 string junction supermultiplet 
by quantizing the zero modes about the monopole of Fig. 2. We will work in
the limit that the brane M approaches straight line joining A and B in 
Figs. 1 and 2. We will find that the expected supermultiplet structure 
emerges only upon assigning bosonic zero modes spin half, and fermionic
zero modes spin zero, in agreement with the analysis of Section 3. 
In this appendix we closely follow the arguments of Section 6 of  \LeeNV.

In the limit that the brane M approaches the straight line joining A and B
in Fig. 2, the monopole \mono\ is marginally unstable to decay into two 
separate monopoles (in pictures, the D1-brane of Fig. 1 can split into two D strings, 
one from A to M, and the second from M to B). At this special point the monopole has 
8 bosonic zero modes, which may be divided into
\item{1.} The centre of mass motions, which consist of the 3 centre
of mass translations of the two monopoles, and a rotation of the overall
$U(1)$.
\item{2.} The relative motions which consist of the 
3 relative separations of the two monopoles, and the rotation in the
relative $U(1)$.
 
Quantization of the $U(1)$ rotations leads to states labeled by two electric
charge integers $\{ n_1, n_2\}$, where $n_1$ and $n_2$ are the number of
quanta of flux travelling from M to A and from M to B respectively, in Fig. 1.

The   bosonic zero modes have corresponding fermionic counterparts. 
The superpartners of the centre of mass motions are the goldstinos 
of the 8 supersymmetries broken by the monopole background. Quantizing these
zero modes yields  a $\CN=4$ vector multiplet, with $2^4$ states. 
The remaining four fermionic zero modes create states with electric charge
$\{\pm 1, 0\}$, $\{0, \pm 1\}$. 

Consider a 3 string junction, as in Fig. 1, with $k$ units of string charge
leaving M. Such a junction may be created by exciting $n_1$ zero mode
creation operators of charge $\{1,0\}$ and $n_2$ zero mode creation operators
of charge $\{0,1\}$ such that 
$n_1+n_2=k$. It is easy to see that the number
of such states is 
\item{1.} $k+1$ states built purely out of bosonic zero modes. 
\item{2.} $2k$ states with 1 fermionic and $k-1$ bosonic quanta. 
\item{3.} $k-1$ states built from two fermionic and $k-2$ bosonic zero modes.  
 
\noindent Tensoring this particle content with the  $\CN=4$ vector
multiplet yields a `multiplet' with  $2^6 k$ particles.

We expect these states to fill out a $({1 \over 4})^{\rm th}$ 
BPS multiplet of 
$\CN=4$ supersymmetry. This expectation is borne out if one
assigns each bosonic zero mode spin half, and each fermionic zero mode spin
zero.  According to this assignment, the states described above constitute 
\item{1.} A single spin ${k \over 2}$ multiplet.
\item{2.} Two spin ${k-1 \over 2}$ multiplets.
\item{3.} A spin ${k-2 \over 2}$ multiplet.

\noindent Tensoring with the  $\CN=4$ vector multiplet yields the particle  content
of a $({1 \over 4})^{\rm th}$ BPS multiplet with maximum spin ${k\over 2}
+1$. This irreducible representation of the supersymmetry algebra 
may be identified with that constructed in \StrathdeeJR\ by tensoring the
basic $({1 \over 4})^{\rm th}$ BPS multiplet (obtained by quantizing the
goldstinos of the 12 broken supercharges) with a spin ${k-1 \over 2}$
representation of $SU(2)$.

\listrefs
\end